# A Metaheuristic Approach for IT Projects Portfolio Optimization

Shashank Pushkar, Abhijit Mustafi and Akhileshwar Mishra

**Abstract**— Optimal selection of interdependent IT Projects for implementation in multi periods has been challenging in the framework of real option valuation. This paper presents a mathematical optimization model for multi-stage portfolio of IT projects. The model optimizes the value of the portfolio within a given budgetary and sequencing constraints for each period. These sequencing constraints are due to time wise interdependencies among projects. A Metaheuristic approach is well suited for solving this kind of a problem definition and in this paper a genetic algorithm model has been proposed for the solution. This optimization model and solution approach can help IT managers taking optimal funding decision for projects prioritization in multiple sequential periods. The model also gives flexibility to the managers to generate alternative portfolio by changing the maximum and minimum number of projects to be implemented in each sequential period.

.**Index Terms**— IT Projects portfolio Management, optimization, Financial Evaluation, Genetic Algorithm , Real Option.

——————————— ◆ ———————————

## 1 INTRODUCTION

IT Projects investment decisions are crucial for any firm to implement e-Business. Widely used Discounted Cash Flow (DCF) method is not appropriate to evaluate IT investments as it is designed for the projects with no option features ([1],[2]). The inability of Discounted Cash Flow analysis to take care of impact of flexibility underlying IT investment decisions have forced IT managers to rely on their gut instinct.[3].

Real Option analysis has been an alternative approach that incorporates impact of flexibility while evaluating IT projects. This project evaluation method has been used based on quantification of projects benefits and costs [4] as well as the risk and volatility of cash flows ([5],[6]).But very few research has scrutinized the relevance of real option analysis of IT investments for optimizing a portfolio of projects. Dickinson et. al. [7] introduces an optimization model for interdependent technology projects but it is not in the framework of real option valuation. Portfolio of IT projects involve interdependencies which can create multiple options ( [8],[9]).Cobb and channes [10] propose an approach for the Real Option Valuation of the portfolio of real investment projects. Bardhan et.al. [11] proposes an integer programming model to get the optimal sequence for implementation of IT projects.Costa et. al.[12] proposes an approach for evaluating the software project risk.Fang and Chen [13] propose a portfolio selection model for a mixed R&D projects.Liang et. al. (2008)[14] give a framework for the IT investment on the basis of Real Option and Mean – Variance theory perspective.Peters and Verhoef[15] quantifies the yields of risk of IT portfolio. Shashank et.al. [16] proposes a dynamic programming solution to this class of problem,but it works for the smaller portfolio as with increaded number of projects the variable becomes very large because of curse of dimensionality.

This paper gives a simplified mathematical model to optimize an IT projects portfolio where projects can be sequentially interdependent (which is an important characteristic of IT projects). The model dynamically calculates option values for each project due to its dependent projects implemented in the period subsequent to it. Genetic Algorithms have a proven track record in handling such large search field problems. The simple act of crossover and mutation allows the algorithm to search a large search space and converge quickly to an optimal solution. In case of exceptionally large search spaces the algorithm provides the option of being terminated after a fixed number of iterations . This may not provide the best solution in all cases but usually gives a very good approximation of the solution. The proposed solution yields the optimum sequence for implementation of IT projects to have the maximum overall portfolio value (DCF+ROV) across multiple time periods. This can help IT managers taking optimal funding decisions. The rest of the paper is organized as follows.

Section 2 specifies the problem The mathematical model is defined in section 3. A Genetic Algorithm solution is given in section 4 and an illustrative numerical example is given in section 5. Finally we conclude in section 6.

- *Shashank Pushkar is with Birla Institute of Technology, Mesra , Ranchi, India-835215.Phone Number: 91-651-227554(O), 91-9470141130.*
- *A Mustafi is with Birla Institute of Technology, Mesra , Ranchi, India-835215. .Phone Number: 91-651-227554(O), 91-9431382747.*
- *.Akhileshwar Mishra is with National Institute of Technology, Department of Computer Application , Jamshedpur Jharkhand, India-831014.*

## 2  PROBLEM SPECIFICATION

The Problem undertaken comes under the category of IT project valuation and investment decisions for project portfolio management. This problem applies to any firm that decides to implement e-Business or wants to invest in multiple IT projects. Here the collection of IT projects is a portfolio of projects to be implemented in sequential periods. The projects are implemented separately but they have two types of dependency. They are as follows:
1. Total dependency
2. Partial dependency
3. No dependency

Total dependency of project i on project j indicates that the capabilities developed for project j is also required by project i i.e. j creates an option to implement i and I can be implemented either together with j or after the implementation of j but not before j's implementation. If i is implemented after implementation of j, Call Option value of i should be added to the DCF value of project j(according to real option analysis).

Partial dependency of project i on j indicates that the capabilities developed for project j supports or enhances the capabilities required by project i, but there is no strict requirement of project i to be implemented together with j or after j's implementation. But if i is implemented without implementation of j, the benefit level of i would reduce by some fraction depending on the level of dependency.

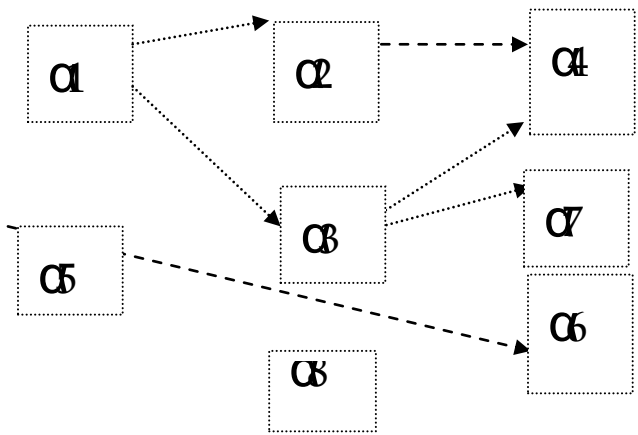

Fig.1 Possible dependencies among the IT projects

In the fig. 1  c1, c2 .etc. .are all investment projects. The solid arrow indicates total dependency (e.g. c2 is totally dependent on c1). Dashed arrow indicates partial dependency (e.g. c6 is partially dependent on c5). The project that has no dependency on any other projects are independent and without arrow.

All these IT projects in the portfolio could not be implemented simultaneously due to the constraints of the budget of the firm and uncertainty regarding success of IT projects (customer response, implementation success within the firm etc.).Therefore these projects are generally implemented stage-wise in sequential periods. Each period may have different budget. The projects implemented in initial stage or period provides opportunities (option) for remaining unimplemented projects which are dependent on implemented ones. Thus the initial implementation of projects provide flexibility to the managers to decide about whether to implement the remaining projects in the portfolio (seeing the response from the customers and the success of projects implementation within the firm ).Since these projects generally share technology and the firm , they share risk. So, the projects implemented in early stages (periods) relax the uncertainties for the remaining unimplemented projects.

According to real option valuation methodologies, projects create option for dependent projects to be implemented in subsequent periods and hence option value of dependent projects must be added as Real Option value to it. So, the total value of a project will be its DCF value plus Real Option values due to its dependent projects to be implemented in subsequent periods. Here it is important to note that the Real Option Value for these dependent projects will not be added to Option generating project which are either implemented before it or together with it.

The problem requires to be formulated as a mathematical model for the optimization of such IT projects portfolio across multiple sequential periods so that the overall value of the portfolio is maximized.

## 3  MATHEMATICAL MODEL

### 3.1 Notation

$p$ = Set of given projects

$n_p$ = Number of project.

$N$ = Total number of periods for completing the implementation of the projects

$B_k$ = Budget for period k (k=1,2,..,N)

$r$ = rate of interest

$C_{ik}$ = Present Value of cost of project i if i is implemented in period k
 ( i = 1,2,..,$n_p$ ; k = 1,2,..,N).
 = cost of project i funded in period $k /(1+ )^{k-1}$
 ( here if k =1, it means the project is funded in the beginning of first period)

$D_i$ = Set of projects directly dependent on project i
 $D_i \subset$ ,

$d_{ij}$ = Level of dependence of project i on project j, defined as follows:
 $d_{ij}$ = , if j is not directly dependent on i;
 $0 < d_{ij} <$ , j is partially dependent on i;
 $d_{ij}$ = , , if j is totally dependent on i.

$R_{ik}$ = Present value of return from project i , if it is implemented in period k.

$$= \sum_{t=?} \cdot /(1+\cdot)^t$$

where,

$r_{it}$ = return from project i at the end of $t^{th}$ period.

$x$ = the expected number of periods up to which the project is going to give the return.

$k$ = period of implementation of project i

$X_{ik}$ = Binary value indicating implementation period for project i, where,

$X_{ik}$ = 1, if i is implemented in period k;
0, otherwise.

and, $\sum_{k=}^{N} X_{ik} =$

i.e. each project i is implemented only once.

$W_{ik}$ = Discounted cash flow(DCF) value of project i if i is implemented in period k.

$V_{ik}$ = The Net Present Option value for project i attributed to its dependent projects if i is implemented in period k.

$V_{ij}$ = Option value for project i due to project j, where

$V_{ij} =$ ⎮ if $j \notin )_i \notin D_i$

$Y_{ijk} =$ , if i is implemented prior to implementation of j
0, otherwise.

$Qk_{min}$ = minimum number of projects to be implemented in period k

$Qk_{max}$ = Maximum number of periods to be implemented in period k

### 3.2 Problem Statement

*It follows from notation above that*

$$V_{ik} = \sum_{j=} j * d_{ij} * X_{ik} * (1 - \sum_{k=} \bar{}_{ik})(i \in \;, j \in \;_i)$$

$$W_{ik} = R_{ik} - \bar{}_{ik}) * X_{ik}, (i \in \;)$$

*The problem is to maximize*

$$\sum_{i \in} \sum_{k=} \bar{}_{ik} + V_{ik})$$

*Subject to the constraints*

1) $\Sigma_i C_{ik} . X_{ik} - B_k \leq 0$ ( Budget Constraint)

2) $Qk_{min} \leq \sum_i X_{ik} \leq Qk_{max}$ ( *Number of projects to be implemented in a period*)

Here the objective function indicates that the real option value of project i due to project j will only be added if i is implemented prior to the implementation of j. Observe that no additional sequencing constraint is required as maximizing the above objective function will automatically take care of the sequencing.

## 4 GENETIC ALGORITHM BASED SOLUTION

Genetic algorithms[18] are meta-heuristic optimization techniques based on natural theory and survival of the fittest. The operators involved in GA tend to be heavily inspired by natural selection and consequently successive generations of the algorithm continue to propagate the best traits of the population. This leads to rapid convergence of the search[25]. Also the introduction of the mutation operator ensures that diversity is not neglected and the search is not trapped in a local maximum. A flow chart illustrating the basics steps of GA based optimization is given in fig 2.

The chromosome structure chosen to represent the problem is a sequence of 'b' bits where

*b =(no of projects\*no of periods)*

A representative bit sequence for a portfolio of seven projects to be completed over three(3) periods would consequently be a 21 bit sequence. It is easy to visualize the above case in terms of integer numbers where any number in the range 0 – 2b-1 would have its binary representation as one of the possible chromosomes.

Once defined the chromosome is divided into N equilength sequences where N is the number of periods under consideration. Every set bit in these subsequences would represent a project to be completed in that particular period.

The GA is then introduced to search over the search space to optimize the NPV of the portfolio

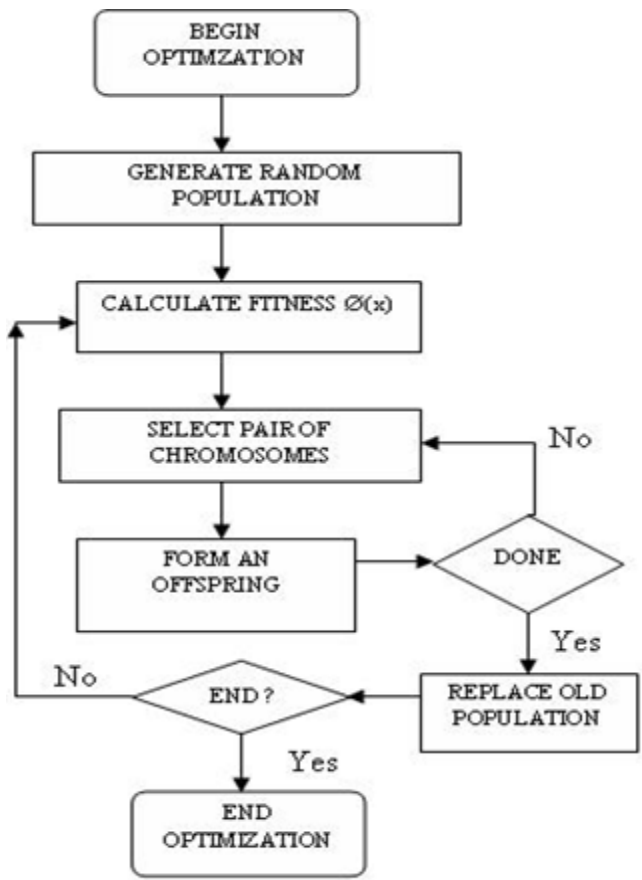

Fig. 2 Flowchart for GA based Optimization

## 5 NUMERICAL EXAMPLE

A seven IT projects portfolio is taken as an illustrative example. The interdependencies among projects are shown in Figure 1 in Section 2 of the paper. The results were simulated using MATLAB® genetic algorithm tollbox.

Here the project planning horizon is taken as N= 3 periods. i.e. k = 1,2,3 and n p = 7.

Other data about the portfolio is listed in the tables below.

TABLE 1- COST-RETURN OF PROJECTS

| Project i | Present Value of cost in period k | Present Value of return in period k |
|---|---|---|
| $c_1$ | 15 | 13 |
| $c_2$ | 30 | 35 |
| $c_3$ | 70 | 65 |
| $c_4$ | 60 | 100 |
| $c_5$ | 15 | 20 |
| $c_6$ | 50 | 150 |
| $c_7$ | 125 | 150 |

∗ Note: Present value of costs and returns for projects to be implemented in period k is taken same for k= 1, 2, 3 for the shake of simplicity.

Table 2 – DEPENDENCY LEVEL AMONG PROJECTS

| | $c_1$ | $c_2$ | $c_3$ | $c_4$ | $c_5$ | $c_6$ | $c_7$ |
|---|---|---|---|---|---|---|---|
| $c_1$ |   | 0 | 0 | 0 | 0 | 0 | 0 |
| $c_2$ | 1 |   | 0 | 0 | 0 | 0 | 0 |
| $c_3$ | 1 | 0 |   | 0 | 0 | 0 | 0 |
| $c_4$ | 0 | 0.25 | 1 |   | 0 | 0 | 0 |
| $c_5$ | 0 | 0 | 0 | 0 |   | 0 | 0 |
| $c_6$ | 0 | 0 | 0 | 0 | 0.25 |   | 0 |
| $c_7$ | 0 | 0 | 1 | 0 | 0 | 0 |   |

TABLE 3 – OPTION VALUES DUE TO DEPENDENCY AMONG PROJECTS

| | $c_1$ | $c_2$ | $c_3$ | $c_4$ | $c_5$ | $c_6$ | $c_7$ |
|---|---|---|---|---|---|---|---|
| $c_1$ |   | 10 | 10 | 0 | 0 | 0 | 0 |
| $c_2$ | 0 |   | 0 | 5 | 0 | 0 | 0 |
| $c_3$ | 0 | 0 |   | 10 | 0 | 0 | 10 |
| $c_4$ | 0 | 0 | 0 |   | 0 | 0 | 0 |
| $c_5$ | 0 | 0 | 0 | 0 |   | 5 | 0 |
| $c_6$ | 0 | 0 | 0 | 0 | 0 |   | 0 |
| $c_7$ | *0* | *0* | *0* | *0* | *0* | *0* |   |

∗Note: Option values for dependent projects are taken same for simplicity but they can be calculated using nested option model option model( Benaroach et.al. [20]) for each project due to its each dependent project.

Budgets for each three periods are as under:
B1 = 90 , B2 = 125, B3 = 175.
$Q_{kmin}$ = 2 and $Q_{kmax}$ = 3 for each k = 1, 2, 3.

TABLE 4-OPTIMIZATION RESULTS

| | K=1 | K=2 | K=3 |
|---|---|---|---|
| Selected Projects for funding | $c_1$, $c_3$ | $c_2$, $c_4$, $c_5$ | $c_6$, $c_7$ |
| Costs of projects | 85 | 105 | 175 |
| Budget | 90 | 125 | 175 |
| PORTFOLIO VALUE | 616.75 | | |

The GA solution given in section IV of the paper is applied to the above example portfolio. The results obtained are tabulated above in Table 4 and the convergence of the algorithm is shown in Fig. below.

The results show that those projects that provide infrastructure to many other projects(thus having high option values) are selected in early periods for funding and those having no option value are deferred to the later periods. This also indicates that the maximization of the option component of portfolio value selects the projects ,which have maximum number of dependent projects ,earlier than those projects which are having less dependent projects or no dependent. Since the projects of these types of portfolio shares risk for being successful, early implementation of high op-tion projects would lower the overall risk of success for the portfolio.

The model yields the optimum value for the overall portfolio along with the periods for the funding of projects in the portfolio. The method calculates the option values of each project due to its dependent projects dynamically and thus represents a significant improvement over the existing models for prioritization of IT projects in real option valuation framework .This approach can help IT managers taking optimal funding decisions .

As a final demonstration the best chromosome for the representative case used in the paper is presented below. The chromosome represents 3 periods of activity and each column has the set bits representing the projects to be completed ideally in that particular period. The solution clearly demonstrates the findings of Table 4.

**100**
**010**
**100**
**010**
**010**
**001**
**001**

Fig. 3 The best chromosome for the case study

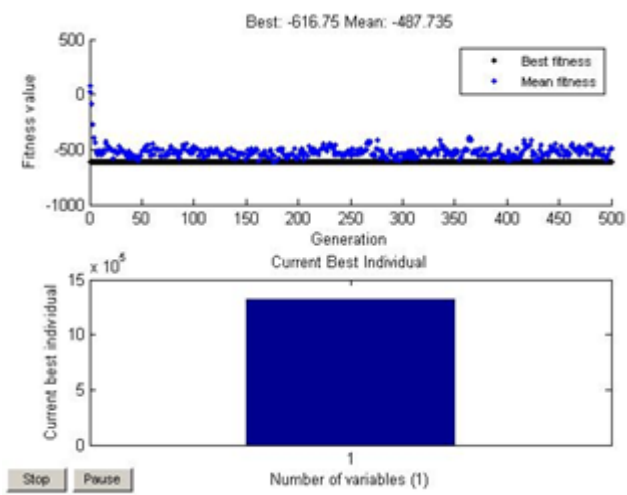

Fig.4 The Convergence and best solution for the GA

## 6 CONCLUDING REMARK

We have developed a multi period portfolio optimization model. The model uses GA to calculate option values for each project due to its dependent projects implemented subsequently. This is the improvement over the existing works where the option values are calculated statically for each projects due to its dependent projects. and not dynamically at the time of implementation decision making. The meta heuristic nature of the solution yields the optimal sequence of implementation of IT projects in multiple periods to get the maximum overall portfolio value .The proposed algorithm is very suitable to solve this problem as it is modeled as a multistage optimization problem and it makes possible to calculate option values for each project due to its subsequently implemented dependent projects to maximize the overall portfolio value. The model also gives flexibility to the managers to generate alternative portfolio by changing the maximum and minimum number of projects to be funded in a period.

The research work can be further extended by incorporating fuzziness in the model as the terms like dependency level among projects, level of benefits etc. are uncertain and may change with the changing decision time

**Shashank Pushkar** is a Lecturer in the Department of Computer Science and Engineering, Birla Institute of Technology , Mesra , Ranchi. His research interest is in the field of Information Technology project Management and Optimization Technique.

**Abhijit Mustafi** is a MCA from the University of North Bengal, India. He is currently a Senior Lecturer in the Department of CSE, BIT mesra, India. His research interests include image processing, meta heuristic algorithms and web mining.

**Dr. Akhileshwar Mishra** is PhD from IIT Kharagpur. He is also a professor of Computer Applications , National Institute of Technology, Jamshedpur. He specializes in computer applications and Optimizations in the field of Industrial Management